\documentclass[12pt,english]{article}
\usepackage[T1]{fontenc}
\usepackage[latin9]{inputenc}
\usepackage[letterpaper]{geometry}
\usepackage{amsmath}
\usepackage{graphicx}
\usepackage{setspace}
\usepackage{amssymb}
\usepackage{verbatim}

\makeatletter

\newcommand{\lanln}[1]{$\langle$\texttt{arXiv:#1}$\rangle$}

\newcommand{\lyxaddress}[1]{
\par {\raggedright #1
\vspace{1.4em}
\noindent\par}
}

\makeatother

\makeatother

\usepackage{babel}

\begin{document}

\title{\begin{flushright}
 \small MZ-TH/10-21
\end{flushright}
\vspace{2cm}
Low energy Quantum Gravity from the \\Effective Average Action}

\author{A. Satz$^{a}\footnote{\texttt{satz@df.uba.ar}}$, A. Codello$^{b}\footnote{\texttt{codello@thep.physik.uni-mainz.de}}$ and F.D. Mazzitelli$^{a}\footnote{\texttt{fmazzi@df.uba.ar}}$}

\maketitle

\lyxaddress{\begin{center}
$^{a}$Departamento de F\'isica, Facultad de Ciencias Exactas y Naturales,\\
Univ. de Buenos Aires
and Instituto de F\'\i sica de Buenos Aires, CONICET\\
Ciudad Universitaria,  Pabellon 1, 1428 Buenos
Aires, Argentina\\
$^{b}$Institut f\"ur Physik, Johannes Gutenberg-Universit\"at,
Mainz\\
Staudingerweg 7, D-55099 Mainz, Germany
\par\end{center}}
\begin{abstract}
Within the effective average action approach to quantum gravity,
we recover the low energy effective action as  derived in the
effective field theory framework, by studying the flow of possibly
non-local form factors that appear in the curvature
expansion of the effective average action. We restrict to the
one-loop flow where progress can be made with the aid of the
non-local heat kernel expansion. We discuss the possible physical
implications of the scale dependent low energy effective action
through the analysis of the quantum corrections to the
Newtonian potential.
\end{abstract}

\section{Introduction}

General relativity and quantum mechanics are not yet unified in a
coherent theory as we do not have yet a fully successful theory of
quantum gravity. But this does not mean that we lack any kind of
quantum gravitational predictions: at least at low energy, quantum
gravity can be described by an effective field theory based on
metric degrees of freedom, as
was first shown by Donoghue and others \cite{Donoghue:1994dn,Burgess:2003jk}.

At scales much smaller than the characteristic scale, which is
here the Planck mass, effective field theory predictions are
possible and calculable. Examples of this kind are the calculation
of the first quantum corrections to the gravitational interaction
potential between two masses \cite{Donoghue:1994dn,BjerrumBohr:2002kt}, and the low-energy graviton scattering cross-section \cite{Donoghue:1999qh}.

The important point about these predictions is that no matter
which theory actually describes high-energy quantum gravity - a
string theory, a spin foam model, or other
approaches - in the infrared (IR) limit any physically valid
theory must reproduce the results
found in the effective field theory framework.

In the last years, the hypothesis that the high energy completion
of quantum gravity can still be described using the metric as
fundamental degrees of freedom has gained some new support. In
particular the possibility that there exists a non-trivial
ultraviolet (UV) fixed point of the renormalization group (RG)
flow, with a finite dimensional UV critical surface, has been
investigated within what is called the {}``Asymptotic Safety''
scenario
\cite{weinberg,Reuter:1996cp,livingrev,Reuter:2007rv,Codello:2008vh}.
Progress in this direction has been possible thanks to the
development of a powerful functional RG framework based on the
construction of a scale dependent effective action, called
effective average action
\cite{Wetterich:1993uk,Reuter:1993kw,Berges:2000ew}. For a general discussion of the relation of this approach to perturbative theory, see \cite{pawlowski}.

In this paper we show how the low energy effective field theory
predictions naturally arise in the effective average action
approach to quantum gravity. In this way we start to delineate a
picture able to describe gravitational phenomena at all scales:
from the UV physics of the non-trivial fixed point down to the IR
physics of the low energy
effective action. 

The basic technical tool at our disposal is the exact RG flow
equation that the effective average action satisfy
\cite{Wetterich:1993uk}. The solution of this integro-differential
equation poses challenging technical difficulties. As of now there
are no completely adequate techniques able to fully extract all
the non-perturbative information which is in principle contained
in the RG flow of the effective average action. In this work we
will focus on a truncation ansatz where we consider an expansion
of the effective average action in powers of the curvature and
where the scale dependence is encoded in the RG running of
(possibly nonlocal) form factors. In this way we project
the flow from the full theory space onto a subspace which is still
infinite dimensional: the functional space of the running
form factors. It was shown in \cite{Codello:2010mj} that to
recover sensible results in the IR limit this is the minimal type
of truncation ansatz we have to consider. Also, we will consider
only the one-loop flow, driven by the Einstein-Hilbert operator
$\int\sqrt{g}R$, of the effective average action, where progress
can be made with the aid of the non-local
heat kernel expansion \cite{Codello:2010mj}.

The paper is organized as follows. In Section 2 we shortly review the
construction of the effective average action for quantum gravity,
we introduce the truncation ansatz we will consider, and we show
how the one-loop flow equations in $d$-dimensions are obtained using the
non-local heat kernel expansion exposed in Appendix A. In
Section 3 we integrate the flow equations for
the form factors in $d=4$ from the ultraviolet (UV) scale to the
infrared (IR) scale recovering in this limit the effective field
theory results. In Section 4 we discuss how the
quantum corrections to the Newtonian potential arise in our
approach. Finally, Section 5 contains our concluding remarks. In the Appendices we
give an overview of the non-local heat kernel expansion and include some details of the calculations.

\section{Effective average action in quantum gravity: curvature expansion}

The effective average action is a coarse-grained effective action
depending on the IR cutoff scale $k$. It interpolates smoothly
between the bare action in the UV, for $k\rightarrow\Lambda$%
\footnote{Here $\Lambda$ is the UV scale, not to be confused with the running
cosmological constant $\Lambda_{k}$.%
}, and the full effective action in the IR, for $k\rightarrow0$
\cite{Wetterich:1993uk,Reuter:1993kw,Berges:2000ew}. In the gravitational context it is
constructed using the background field formalism \cite{Reuter:1996cp,Reuter:2007rv}, and
is a functional $\Gamma_{k}[h,\bar{C},C;\bar{g}]$ depending on the
average metric fluctuation $h_{\mu\nu}$, on the average ghost
fields $\bar{C}_{\mu},C^{\mu}$ and on the background metric
$\bar{g}_{\mu\nu}$. The full quantum field is recovered in the sum
$g_{\mu\nu}=\bar{g}_{\mu\nu}+h_{\mu\nu}$. The background metric is used to implement in a
gauge invariant way the coarse-graining procedure. A cutoff
operator constructed with the background metric - generally, a
Laplacian type operator - is used to divide the slow field modes
from the fast field modes, which are then integrated out.\\
Actually, $\Gamma_{k}[h,\bar{C},C;\bar{g}]$ is invariant only
under combined physical plus background gauge transformations; in
its most general form can be written as
\begin{equation}
\Gamma_{k}[h,\bar{C},C;\bar{g}]=\bar{\Gamma}_{k}[\bar{g}+h]+\hat{\Gamma}_{k}[h,\bar{C},C;\bar{g}]\,,\label{1}
\end{equation}
where the functionals introduced are defined by the relations
$\hat{\Gamma}_{k}[0,0,0;\bar{g}]=0$ and
$\bar{\Gamma}_{k}[\bar{g}]=\Gamma_{k}[0,0,0;\bar{g}]$. It follows
that the functional $\bar{\Gamma}_{k}[g]$ is invariant under
physical gauge transformations while the functional
$\hat{\Gamma}_{k}[h,\bar{C},C;\bar{g}]$ is invariant under
physical plus background gauge transformations and can be
interpreted as a generalized background gauge-fixing and ghost
action. In this paper we will restrict ourselves to a commonly
used truncation where $\hat{\Gamma}_{k}[h,\bar{C},C;\bar{g}]$ is
approximated by the classical background gauge-fixing and ghost
actions:
\begin{equation}
\hat{\Gamma}_{k}[h,\bar{C},C;\bar{g}]=S_{gf}[h;\bar{g}]+S_{gh}[h,\bar{C},C;\bar{g}]\,.\label{2}
\end{equation}
Here the background gauge-fixing condition is
$f_{\mu}[h,\bar{g}]=0$, with\[
f_{\mu}[h,\bar{g}]=\left(\delta_{\mu}^{\alpha}\bar{\nabla}^{\beta}-\frac{\beta}{2}\bar{g}^{\alpha\beta}\bar{\nabla}_{\mu}\right)h_{\alpha\beta}\,;\]
the background gauge-fixing action is
\begin{equation}
S_{gf}[h;\bar{g}]=\frac{1}{2\alpha}\int
\mathrm{d}^{d}x\sqrt{\bar{g}}\bar{g}^{\mu\nu}f_{\mu}[h,\bar{g}]f_{\nu}[h,\bar{g}]\,.
\label{gf}
\end{equation}
$\alpha$ and $\beta$ are gauge-fixing parameters that we will fix
to $\alpha=\beta=1$. The ghost action related to the background
gauge-fixing condition is\[ S_{gh}[h,\bar{C},C;\bar{g}]=-\int
\mathrm{d}^{d}x\sqrt{\bar{g}}\,\bar{C}^{\mu}\left(\bar{\nabla}^{\alpha}g_{\nu\alpha}\nabla_{\mu}+\bar{\nabla}^{\alpha}g_{\mu\nu}\nabla_{\alpha}-\beta\bar{\nabla}_{\mu}g_{\nu\alpha}\nabla^{\alpha}\right)C^{\nu}\,.\]

With this choice for $\hat{\Gamma}_{k}[\varphi;\bar{g}]$ all the
scale dependence is in the functional $\bar{\Gamma}_{k}[g]$.
Usually (e.g. \cite{Reuter:2007rv,Codello:2008vh}) truncations of $\bar{\Gamma}_{k}[g]$ have been chosen to
be a sum of local operators with $k$-dependent coupling constants
as coefficients. Truncations involving functions of the Ricci
scalar as integrand have been considered in \cite{Codello:2007bd,Machado:2007ea}. Here we
introduce a general truncation scheme, which may be called
{}``curvature expansion'', where we expand $\bar{\Gamma}_{k}[g]$
in powers of the curvatures, retaining all possible terms involving
the D'Alambertian operator $\square=\nabla_{\mu}\nabla^{\mu}$ by the way of form factors. To order $\mathcal{R}^{2}$, we consider\footnote{$\mathcal{R}$ being any curvature.%
}:
\begin{equation} \bar{\Gamma}_{k}[g]=\int
\mathrm{d}^{d}x\sqrt{g}\left[\frac{1}{16\pi
G_{k}}\left(2\Lambda_{k}-R\right)+RF_{1,k}\left(-\square\right)R+R_{\mu\nu}F_{2,k}\left(-\square\right)R^{\mu\nu}\right]+O\left(\mathcal{R}^{3}\right)\,.\label{2.51}
\end{equation}
This is the kind of general action that is postulated in effective
field theory in a regime where
$\mathcal{R}^{2}\ll\nabla\nabla\mathcal{R}$ (see \cite{Barvinsky:1990up, Vilkovisky:gospel});
it is, in fact, the most general action quadratic in the
curvature. Note that the form factors $F_{i,k}(x)$
can make the effective average action (\ref{2.51}) non-local.

It is possible to derive an exact RG flow equation describing the
dependence of $\bar{\Gamma}_{k}[g]$ on the RG parameter $t=\log
k/k_{0}$, where $k_{0}$ is an arbitrary reference scale. This
reads \cite{Reuter:1996cp}:
\begin{equation}
\partial_{t}\bar{\Gamma}_{k}[g]=\frac{1}{2}\textrm{Tr}_{2}\frac{\partial_{t}R_{k}(\Delta_{2})}{\Gamma_{k}^{(2,0,0;0)}[0,0,0;g]+R_{k}(\Delta_{2})}-\textrm{Tr}_{1}\frac{\partial_{t}R_{k}(\Delta_{1})}{\Gamma_{k}^{(0,1,1;0)}[0,0,0;g]+R_{k}(\Delta_{1})}\,,\label{2.1}
\end{equation}
$R_{k}(\Delta_{2})$ and $R_{k}(\Delta_{1})$ are, respectively, the
cutoff kernels of spin-two gravitons and spin-one ghosts; they are
functions of the cutoff operators $\Delta_{2}$ and $\Delta_{1}$
(to be chosen in a moment) used to separate the slow field modes
from the fast ones. The functional form $R_{k}(z)$ is arbitrary
except for the requirements that it should be a monotonically
decreasing function in both $z$ and $k$, that
$R_{k}(z)\rightarrow0$ for $z\gg k^{2}$
and that $R_{k}(z)\rightarrow k^{2}$ for $z\ll k^{2}$. The notation $\Gamma_{k}^{(2,0,0;0)}[0,0,0;g]$
and $\Gamma_{k}^{(0,1,1;0)}[0,0,0;g]$ is used for the Hessians of the functional $\Gamma_{k}[h,\bar{C},C;\bar{g}]$.

Since to insert in the flow equation (\ref{2.1}) the full
truncation (\ref{2.51}) is technically a very difficult task, in
this paper we will adopt a one-loop approximation: when computing
the Hessians in the right hand side of (\ref{2.1}), we drop the
contributions from the form factors, we set
$\Lambda_{k}=0$, and we disregard the running of the
wave-functions renormalization of the metric fluctuation and of
the ghosts that are contained in the cutoff kernels. Computing the
traces and integrating the flow will then give us the one-loop
running of the couplings $G_{k},\Lambda_{k}$ and of the form
factors $F_{i,k}(x)$ induced by the operator $\int\sqrt{g}R$ in
the Einstein-Hilbert part of the ansatz (\ref{2.51}). 

Within this approximation the flow equation (\ref{2.1}), after
calculating the Hessians, becomes:
\begin{equation}
\partial_{t}\bar{\Gamma}_{k}[g]=\frac{1}{2}\textrm{Tr}_{2}\frac{\partial_{t}R_{k}(\Delta_{2})}{\Delta_{2}+R_{k}(\Delta_{2})}-\textrm{\textrm{Tr}}_{1}\frac{\partial_{t}R_{k}(\Delta_{1})}{\Delta_{1}+R_{k}(\Delta_{1})}\,.\label{3}
\end{equation}
We have chosen the cutoff operators $\Delta_{2}$ and $\Delta_{1}$
to be the inverse graviton and ghost propagators; this corresponds
to a type II cutoff in the nomenclature of \cite{Codello:2008vh}.
These inverse propagators read:
\begin{eqnarray}
(\Delta_{2})_{\rho\sigma}^{\alpha\beta} & = & -\square\delta_{\rho\sigma}^{\alpha\beta}+R\,\left(\delta_{\rho\sigma}^{\alpha\beta}-\frac{1}{2}g^{\alpha\beta}g_{\rho\sigma}\right)+\left(g^{\alpha\beta}R_{\rho\sigma}+R^{\alpha\beta}g_{\rho\sigma}\right)+\nonumber \\
 &  & -\frac{1}{2}\left(\delta_{\rho}^{\alpha}R_{\sigma}^{\beta}+\delta_{\sigma}^{\alpha}R_{\rho}^{\beta}+R_{\rho}^{\alpha}\delta_{\sigma}^{\beta}+R_{\sigma}^{\alpha}\delta_{\rho}^{\beta}\right)-\left(R_{\;\rho\;\,\;\sigma}^{\beta\;\,\alpha}+R_{\;\sigma\;\,\;\rho}^{\beta\;\,\alpha}\right)+\nonumber \\
 &  & -\frac{d-4}{2(d-2)}\left(R\, g^{\alpha\beta}g_{\rho\sigma}+g^{\alpha\beta}R_{\rho\sigma}+R^{\alpha\beta}g_{\rho\sigma}\right)\,.\nonumber \\
(\Delta_{1})_{\nu}^{\mu} & = &
-\square\delta_{\nu}^{\mu}-R_{\nu}^{\mu}\,.\label{4}
\end{eqnarray}
The running of the zeroth and first order terms in the curvature,
in other words of the running of the cosmological constant
$\Lambda_{k}$ and of Newton's constant $G_{k}$, can be computed
without need of the one-loop approximation. This computation has
been done, within a type II cutoff here considered, in
\cite{Codello:2008vh}, so we will limit ourselves to
noting the one-loop approximation of the result, which
is:
\begin{eqnarray}\label{betas}
\partial_{t}\tilde{\Lambda}_{k} & = & -2\tilde{\Lambda}_{k}+\frac{8\pi(d-3)}{(4\pi)^{d/2}\Gamma\left(\frac{d}{2}\right)}\tilde{G}_{k}\nonumber \\
\partial_{t}\tilde{G}_{k} & = & (d-2)\tilde{G}_{k}-\frac{4\pi\left(5d^{2}-3d+24\right)}{3(4\pi)^{d/2}\Gamma\left(\frac{d}{2}\right)}\tilde{G}_{k}^{2}\,.\label{4.1}
\end{eqnarray}
where $\tilde{\Lambda}_{k}=k^{-2}\Lambda_{k}$ and $\tilde{G}_{k}=k^{d-2}G_{k}$
are the dimensionless couplings. In (\ref{4.1}) the optimized cutoff defined 
later in (\ref{optim}) was used. Already at the one-loop level here
considered, the beta functions (\ref{4.1}) exibits, in $d=4$, a
non-trivial UV attractive fixed-point at the couplings values $\tilde{\Lambda}_{*}=\frac{3}{46}\simeq0.07$
and $\tilde{G}_{*}=\frac{6\pi}{23}\simeq0.82$. For more details on
the Asymptotic Scenario see \cite{Codello:2008vh}.

The novel contribution in this paper lies in the running of the
form factors in $\bar{\Gamma}_{k}[g]$. Their running is obtained
expanding the trace on the right hand side of the flow equation (\ref{3})
in a curvature power series and computing the $O(\mathcal{R}^{2})$
terms. This can be done with the non-local heat kernel expansion methods
detailed in Appendix A. We find:
\begin{eqnarray}
\left.\partial_{t}\bar{\Gamma}_{k}[g]\right|_{\mathcal{R}^{2}} & = & \frac{1}{(4\pi)^{d/2}}\int \mathrm{d}^{d}x\sqrt{g}\left\{ R\left[\int_{0}^{\infty}ds\,\tilde{h}_{k}(s)\, s^{2-\frac{d}{2}}\, f_{1}(-s\square)\right]R \right.\nonumber \\ &  & \left.+R_{\mu\nu}\left[\int_{0}^{\infty}ds\,\tilde{h}_{k}(s)\, s^{2-\frac{d}{2}}\, f_{2}(-s\square)\right]R^{\mu\nu}
\right\} \,.\label{7}
\end{eqnarray}
Here we defined the function $h_{k}(z)=\frac{\partial_{t}R_{k}(z)}{z+R_{k}(z)}$,
of which $\tilde{h}_{k}(s)$ is the anti-Laplace transform. The functions
$f_{1}(x)$ and $f_{2}(x)$ in (\ref{7}) are derived combining the
non-local heat kernel functions for the operators (\ref{4}) and expressing
them in the $\{R^{2},R_{\mu\nu}R^{\mu\nu}\}$ basis, as further explained
in Appendix A; they read:
\begin{eqnarray}
f_{1}(x) & = & \frac{9d^{3}-61d^{2}-10d+320}{128(d-2)}f(x)-\frac{3d^{2}+7d+16}{32x}f(x)\nonumber \\
 &  & +\frac{17d^{2}+45d+96}{192x}-\frac{d(d-3)}{32x^{2}}\left[f(x)-1\right]\,, \nonumber \\
f_{2}(x) & = &\frac{d^{2}+2d-16}{4(d-2)}f(x)+\frac{d}{x}f(x)-\frac{(27-d)d}{24x}+\frac{d(d-3)}{4x^{2}}\left[f(x)-1\right]\,, \label{8}
\end{eqnarray}
where the basic heat kernel non-local form factor $f(x)$ is defined by
\begin{equation}
f(x)=\int_{0}^{1}\mathrm{d}\xi\mathrm{e}^{-\xi(1-\xi)x}\,.\label{fdef}
\end{equation}
From (\ref{7}) we can extract the running of the form factors:
\begin{equation}
\partial_{t}F_{i,k}=\frac{1}{(4\pi)^{d/2}}\int_{0}^{\infty}\mathrm{d}s\,\tilde{h}_{k}(s)\, s^{2-\frac{d}{2}}\, f_{i}(-s\square)\,,\label{9}
\end{equation}
for $i=1,2$. For each $i$ this can be rewritten in terms of a combination
of $Q$-functionals within parameter integrals. Inserting (\ref{8}) in (\ref{9}) and using the definitions of the $Q$-functionals from Appendix B, we find the form:
\begin{eqnarray}
(4\pi)^{d/2}\,\partial_{t}F_{1,k}(x) & = &  \frac{9d^{3}-61d^{2}-10d+320}{128(d-2)}\int_{0}^{1}\mathrm{d}\xi\, Q_{\frac{d}{2}-2}\left[h_{k}\left(z+x\xi(1-\xi)\right)\right]\nonumber \\
 &  & -\frac{3d^{2}+7d+16}{32x}\int_{0}^{1}\mathrm{d}\xi\, Q_{\frac{d}{2}-1}\left[h_{k}\left(z+x\xi(1-\xi)\right)\right]\nonumber \\
 &  & +\frac{17d^{2}+45d+96}{192x}Q_{\frac{d}{2}-1}[h_{k}]\nonumber \\
 &  & -\frac{d(d-3)}{32x^{2}}\left\{ \int_{0}^{1}\mathrm{d}\xi\, Q_{\frac{d}{2}}\left[h_{k}\left(z+x\xi(1-\xi)\right)\right]-Q_{\frac{d}{2}}[h_{k}]\right\} \label{10}
\end{eqnarray}
and
\begin{eqnarray}
(4\pi)^{d/2}\,\partial_{t}F_{2,k}(x) & = & 
\frac{d^{2}+2d-16}{4(d-2)}\int_{0}^{1}\mathrm{d}\xi\, Q_{\frac{d}{2}-2}\left[h_{k}\left(z+x\xi(1-\xi)\right)\right]\nonumber \\
 &  & +\frac{d}{x}\int_{0}^{1}\mathrm{d}\xi\, Q_{\frac{d}{2}-1}\left[h_{k}\left(z+x\xi(1-\xi)\right)\right]-\frac{(27-d)d}{24x}Q_{\frac{d}{2}-1}[h_{k}]\nonumber \\
 &  & +\frac{d(d-3)}{4x^{2}}\left\{ \int_{0}^{1}\mathrm{d}\xi\, Q_{\frac{d}{2}}\left[h_{k}\left(z+x\xi(1-\xi)\right)\right]-Q_{\frac{d}{2}}[h_{k}]\right\}\,,\label{11}\end{eqnarray}
where $x$ stands for $-\square$.

Equations (\ref{10}) and (\ref{11}) are the contributions induced
by the operator $\int\sqrt{g}R$, within a type II cutoff, to the
flow of the curvature squared form factors in truncation (\ref{2.51}).
They are valid in general dimension and for arbitrary cutoff shape
function. These are the main results of this section.

\section{Non-local effective average action in four dimensions: flow and renormalization}

We now study the flow (\ref{10}) and (\ref{11}) of the form
factors in the physical dimension $d=4$. We find:
\begin{eqnarray}
(4\pi)^{2}\,\partial_{t}F_{1,k}(x) & = & -\frac{15}{32}\int_{0}^{1}\mathrm{d}\xi\, Q_{0}\left[h_{k}\left(z+x\xi(1-\xi)\right)\right]-\frac{23}{8x}\int_{0}^{1}\mathrm{d}\xi\, Q_{1}\left[h_{k}\left(z+x\xi(1-\xi)\right)\right]+\nonumber \\
 &  & +\frac{137}{48x}Q_{1}[h_{k}]-\frac{1}{8x^{2}}\left\{ \int_{0}^{1}\mathrm{d}\xi\, Q_{2}\left[h_{k}\left(z+x\xi(1-\xi)\right)\right]-Q_{2}[h_{k}]\right\}\label{12}
\end{eqnarray}
and
\begin{eqnarray}
(4\pi)^{2}\,\partial_{t}F_{2,k}(x) & = & \int_{0}^{1}\mathrm{d}\xi\, Q_{0}\left[h_{k}\left(z+x\xi(1-\xi)\right)\right]+\frac{4}{x}\int_{0}^{1}\mathrm{d}\xi\, Q_{1}\left[h_{k}\left(z+x\xi(1-\xi)\right)\right]+\nonumber \\
 &  & -\frac{23}{6x}Q_{1}[h_{k}]+\frac{1}{x^{2}}\left\{ \int_{0}^{1}\mathrm{d}\xi\, Q_{2}\left[h_{k}\left(z+x\xi(1-\xi)\right)\right]-Q_{2}[h_{k}]\right\} \,.\label{13}
\end{eqnarray}
The previous equations can be rewritten as:
\begin{equation}
\partial_{t}F_{i,k}(x)=\frac{1}{(4\pi)^{2}}g_{i}\left(\frac{x}{k^{2}}\right)\,,\label{14}
\end{equation}
where the functions $g_{i}(u)$ can be calculated once a cutoff shape
function has been chosen. Note that, in $d=4$, all the $k$-dependence
is through the $x/k^{2}$ dependence of the functions $g_{i}(u)$.
If we employ the optimized cutoff shape function \cite{Litim:2001up,litim2}:
\begin{equation}
R_{k}(z)=\left(k^{2}-z\right)\theta(k^{2}-z)\,,\label{optim}
\end{equation}
we find (see the Appendix B for the relevant integrals used):
\begin{eqnarray}
g_{1}(u) & = &\frac{1}{60}+\left(-\frac{1}{60}+\frac{19}{5u}+\frac{1}{15u^{2}}\right)\sqrt{1-\frac{4}{u}}\theta(u-4)\,\label{15}\\
g_{2}(u) & = & \frac{7}{10}-\left(\frac{7}{10}+\frac{76}{15u}+\frac{8}{15u^{2}}\right)\sqrt{1-\frac{4}{u}}\theta(u-4)\,
.\label{16}
\end{eqnarray}
These are the beta functions for the non-local form
factors. They are plotted in Figure 1.%
\begin{figure}
\centering{}\includegraphics{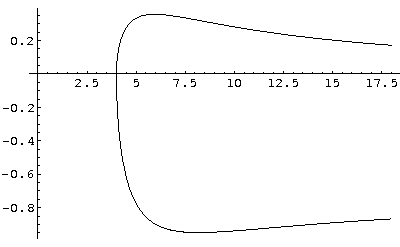} \caption{The
functions $g_{1}(u)-\frac{1}{60}$ (upper curve) and
$g_{2}(u)-\frac{7}{10}$ (lower curve), representing the flow of the form factors in (\ref{14})
after we imposed the UV boundary conditions (\ref{20}), as a function of $u$. 
Note that the flow stops for $x<4k^{2}$: only IR or slow modes contribute effectively
to the RG running of the form factors.}
\end{figure}

The functions (\ref{15}) and (\ref{16}) are constant in a 
neighborhood of the origin: if instead of considering the full functional 
dependence of the form factors $F_{i,k}(x)$ on the D'Alambertian, we had considered
only a local expansion to a polynomial, we would had found zero
beta functions for the running couplings of all derivative terms of the form 
$ \int\sqrt{g}R_{\mu\nu}(-\square)^{n}R^{\mu\nu} $ or $ \int\sqrt{g} R(-\square)^{n}R $. 
As already noticed \cite{Codello:2010mj}, this
shows that truncations to a finite number of local terms
are generally not powerful enough to describe correctly IR physics.
Truncations of the effective average action need at least to project the flow 
onto an infinite dimensional subspace of theory space, as here the one of 
the functions $ F_{i,k}(x)$, to correctly recover the effective action for $k=0$.

We now integrate the flow equations (\ref{14}) from a UV scale
$\Lambda$ down to a generic IR scale $k$. It is expected that for
$\Lambda\rightarrow\infty$ we will encounter the usual ultraviolet
divergences that are found in one-loop effective theory. Using
$\partial_{t}=k\partial_{k}$ in (\ref{14}), we find
\begin{equation}
F_{i,\Lambda}(x)-F_{i,k}(x)=\frac{1}{(4\pi)^{2}}\int_{k}^{\Lambda}\frac{\mathrm{d}k'}{k'}g_{i}\left(\frac{x}{k'^{2}}\right)\,,
\end{equation}
and after going to the variable $y=x/k^{2}$ (with $\mathrm{d}k/k=-\mathrm{d}y/2y$)
we get
\begin{equation}
F_{i,\Lambda}(x)-F_{i,k}(x)=\frac{1}{(4\pi)^{2}}\int_{x/\Lambda^{2}}^{x/k^{2}}\frac{\mathrm{d}y}{2y}g_{i}\left(y\right)\,.\label{17}
\end{equation}
The constant terms in the flow functions (\ref{15}) and (\ref{16})
make the integrals in (\ref{17}) logarithmically divergent at the
lower limit when $\Lambda\rightarrow\infty$. We can isolate this
divergences in the following way:
\begin{eqnarray}
F_{1,\Lambda}(x)-F_{1,k}(x) & = & \frac{1}{(4\pi)^{2}}\frac{1}{60}\left(\log\frac{\Lambda}{k_{0}}+\log\frac{k_{0}}{k}\right)+\frac{1}{(4\pi)^{2}}\int_{x/\Lambda^{2}}^{x/k^{2}}\frac{\mathrm{d}y}{2y}\left[g_{1}\left(y\right)-\frac{1}{60}\right]\label{18}\\
F_{2,\Lambda}(x)-F_{2,k}(x) & = & \frac{1}{(4\pi)^{2}}\frac{7}{10}\left(\log\frac{\Lambda}{k_{0}}+\log\frac{k_{0}}{k}\right)+\frac{1}{(4\pi)^{2}}\int_{x/\Lambda^{2}}^{x/k^{2}}\frac{\mathrm{d}y}{2y}\left[g_{2}\left(y\right)-\frac{7}{10}\right]\,,\label{19}\end{eqnarray}
where $k_{0}$ is an arbitrary reference scale which plays a role
akin to $\mu$ in effective field theory. The logarithmic $\Lambda$
terms in (\ref{18}) and (\ref{19}) correspond to the UV divergences
first calculated in \cite{thooft}. We can renormalize
the theory imposing the UV boundary conditions:
\begin{eqnarray}
F_{1,\Lambda}(x) & = & \frac{1}{(4\pi)^{2}}\frac{1}{60}\log\frac{\Lambda}{k_{0}}+c_{1}\nonumber \\
F_{2,\Lambda}(x) & = &
\frac{1}{(4\pi)^{2}}\frac{7}{10}\log\frac{\Lambda}{k_{0}}+c_{2}\,,\label{20}
\end{eqnarray}
where the $c_{i}$ are possible finite renormalizations\footnote{We stress that there is no nontrivial assumption in setting these boundary conditions: the functions $F_{i,k}$ diverge logarithmically for large $k$ irrespectively of this choice, which is just a technical convenience for eliminating $\Lambda$. This is a one-loop result and does not conflict with previous results about the non-perturbative fixed point for the curvature squared couplings \cite{benedetti}.}. The
important point in the equations (\ref{18}) and (\ref{19}) is that
the integrals are now convergent in the lower limit when we take
$\Lambda\rightarrow\infty$. The scale-dependent form factors at the scale $k$ turn out to be:
\begin{eqnarray}
F_{1,k}(x) & = & \frac{1}{32\pi^{2}}\Bigg\{\Bigg[\frac{1}{30}\log\Bigg(\frac{1+\sqrt{1-\frac{4k^{2}}{x}}}{2}\Bigg)+\left(-\frac{601}{900}+\frac{1139k^{2}}{450x}+\frac{2k^{4}}{75x^{2}}\right)\sqrt{1-\frac{4k^{2}}{x}}\nonumber \\
 &  & +\frac{1}{60}\log\left(\frac{x}{k_{0}^{2}}\right)\Bigg]\theta(x-4k^{2})+\frac{1}{60}\log\left(\frac{k^{2}}{k_{0}^{2}}\right)\theta(4k^{2}-x)\Bigg\}\nonumber \\
F_{2,k}(x) & = &  \frac{1}{32\pi^{2}}\Bigg\{\Bigg[\frac{7}{5}\log\Bigg(\frac{1+\sqrt{1-\frac{4k^{2}}{x}}}{2}\Bigg)-\left(\frac{41}{75}+\frac{84k^{2}}{25x}+\frac{16k^{4}}{75x^{2}}\right)\sqrt{1-\frac{4k^{2}}{x}}\nonumber \\
 &  & +\frac{7}{10}\log\left(\frac{x}{k_{0}^{2}}\right)\Bigg]\theta(x-4k^{2})+\frac{7}{10}\log\left(\frac{k^{2}}{k_{0}^{2}}\right)\theta(4k^{2}-x)\Bigg\}\,.\label{20.1}
\end{eqnarray}
These results are now to be reinserted in the
$O\left(\mathcal{R}^{2}\right)$ part of the truncation ansatz
(\ref{2.51}):
\begin{equation}
\left.\bar{\Gamma}_{k}[g]\right|_{\mathcal{R}^{2}}=\int
\mathrm{d}^{4}x\,\sqrt{g}\left[RF_{1,k}(-\square)R+R_{\mu\nu}F_{2,k}(-\square)R^{\mu\nu}\right]\,.\label{20.2}
\end{equation}
This is the result we were looking for. Note that the form factors are continuous at $x=4k^{2}$, and that for
$k\rightarrow0$ we obtain a well-defined limit, namely the action:
\begin{equation}
\left.\bar{\Gamma}_{0}[g]\right|_{\mathcal{R}^{2}}=\frac{1}{32\pi^{2}}\int
\mathrm{d}^{4}x\sqrt{g}\left[\frac{1}{60}R\,\log\left(\frac{-\square}{k_{0}^{2}}\right)\,
R+\frac{7}{10}R_{\mu\nu}\,\log\left(\frac{-\square}{k_{0}^{2}}\right)\,
R^{\mu\nu}\right] \,,\label{21}\end{equation} where we fixed the finite
renormalizations in (\ref{20}) to
$c_{1}=-\frac{1}{(4\pi)^{2}}\frac{601}{1800}$ and
$c_{2}=-\frac{1}{(4\pi)^{2}}\frac{41}{150}$. The resulting
non-local terms equal the part of standard one-loop quantum
gravity effective action that arises solely from graviton and
ghosts vacuum polarization \cite{Dalvit:1997yc}.

In summary, we have obtained a non-local effective average action, equations (\ref{20.1}) and (\ref{20.2}),
that flows from an ultraviolet scale $k=\Lambda$ to the infrared
limit $k=0$, and in the latter limit equals the expected effective field theory result.
In the next section we will discuss possible physical effects arising from this effective average action.

\section{Quantum Newtonian potential from the effective average action}
In this section we will compute the quantum corrections to the
Newtonian potential stemming from our effective action.
We couple the gravity effective action to a classical matter source:
\begin{eqnarray}
\bar{\Gamma}_k[g] &=& \frac{1}{16\pi G_k}\int \mathrm{d}^4
x\,\sqrt{g}\left(2\Lambda_k-R\right)+\int \mathrm{d}^4
x\,\sqrt{g}\left[R F_{1,k}(-\square)R+R_{\mu\nu}
F_{2,k}(\square)
R^{\mu\nu}\right]\nonumber\\
&+& \int\sqrt{g}\,\mathcal{L}_{\mathrm{mat}}\,.
\end{eqnarray}
Note that the matter action is taken to be scale independent.
From this action we will derive the equations of motion under the
assumption that the gravitational field is weak throughout space
in addition to static.
Before presenting the explicit calculations, we would like to clarify
two important points. Firstly, we will switch to 
work in a 3+1 static spacetime,  analytically continuing our
Euclidean expressions for the form factors into the Lorentzian
sector. With the usual definition of the ``in-out'' effective action, this would imply a replacement
of the Euclidean propagators by the corresponding Feynman propagators in the form factors, and the resulting
field equations would be neither real nor causal. In order to get real and causal equations,
one can introduce a ``Closed Time Path'' (CTP) or ``in-in'' effective action \cite{calzettahu}.
As shown in \cite{Barvinsky:1987}, when the quantum fluctuations are in the vacuum state,
the CTP procedure is equivalent to the replacement of the Euclidean by the retarded propagators
in the form factors appearing in the  field equations. Due to the staticity assumption, this is equivalent to
the replacement $\square\rightarrow \nabla^2$ (the 3-Laplacian) in the form factors.
Secondly, as pointed out in \cite{Dalvit:1997yc}, the solutions of the field equations derived
from the effective action will depend on the gauge fixing parameters, and therefore they
are not physical. Indeed, for our calculations we considered the particular values $\alpha=\beta=1$ in the gauge fixing
condition (\ref{gf}). In the general case, the effective action and the quantum corrections
to the metric will depend explicitly on $\alpha$ and $\beta$. In order to obtain physical results, it is necessary
to define an observable
from the quantum corrected metric, as proposed in \cite{Dalvit:1997yc}. Though important, this issue will not be
relevant in the discussion that follows.

We write $g_{\mu\nu}(\mathbf{x}) = \eta_{\mu\nu} +
h_{\mu\nu}(\mathbf{x})$ with $h_{\mu\nu}$ small everywhere, and  we
will proceed assuming $k$ to be a fixed parameter, ignoring for
the moment the possibility (discussed e.g. in \cite{Bonanno:2000ep,Reuter:2003ca,Reuter:2004nv,Falls:2010he}, and also later in this paper) that it should
depend on position.
The equations of motion are:
\begin{equation}
R_{\mu\nu}-\frac{1}{2}R\,g_{\mu\nu}-\Lambda_k g_{\mu\nu}=8\pi G_k
 \left( T_{\mu\nu}^{\mathrm{mat}}- F_{1,k}(-\nabla^2) H^{(1)}_{\mu\nu} - F_{2,k}(-\nabla^2)H^{(2)}_{\mu\nu} \right)
\end{equation}
where $H^{(1)}_{\mu\nu}=4\nabla_\mu\nabla_\nu
R-4\eta_{\mu\nu}\nabla^2 R$ and
$H^{(2)}_{\mu\nu}=2\nabla_\mu\nabla_\nu R-\eta_{\mu\nu}\nabla^2
R-2\nabla^2 R_{\mu\nu}$ are the variations of the squared
curvature scalar and squared Ricci tensor. All curvature tensors are evaluated at first order
in $h_{\mu\nu}$ . We will assume in what follows that we are in a
non-cosmological regime where $\Lambda_k$ can be neglected.

We choose the classical matter to be static
and nonrelativistic so that there exists a quasi-Cartesian coordinate system in which 
\begin{equation}
T_{\mu\nu}^{\mathrm{mat}}=
\mathrm{diag}(\rho(\mathbf{x}),0,0,0)\, .
\end{equation}
We write for the metric
perturbation $h_{\mu\nu} = h^{c}_{\mu\nu} + h^{q}_{\mu\nu}$ where
$h^{c}_{\mu\nu}$ solves the classical equations of motion and
$h^{q}_{\mu\nu}$ is $O(\hbar)$; also, we write $G_k= G_0(1+\delta
G_k)$ with $G_0$ being the experimental value of $G$, assumed to
be measured at $k=0$, and $\delta G_k$ being $O(\hbar)$ .
Expanding to the first order in $\hbar$ and working in the Lorentz
gauge, the equations for the classical and quantum parts of the
metric read:
\begin{subequations}
\begin{align}
\nabla^2 h_{\mu\nu}^c &= -16\pi G_0 (T_{\mu\nu}^{\mathrm{mat}}-\frac{1}{2}\eta_{\mu\nu}\eta^{\lambda\kappa}T_{\lambda\kappa}^{\mathrm{mat}})= -8\pi G_0\, \rho\, \mathrm{diag}(1,1,1,1)  \label{poisson}\\
\nabla^2 h_{\mu\nu}^q &= 16\pi G_0
\Big[\left(4F_{1,k}(-\nabla^2)+2F_{2,k}(-\nabla^2)\right)\partial_\mu\partial_\nu
R+\eta_{\mu\nu}\left(2F_{1,k}(-\nabla^2)+F_{2,k}(-\nabla^2)\right)\nabla^2R\nonumber\\
&-2F_{2,k}(-\nabla^2)\nabla^2R_{\mu\nu}
-\frac{1}{2}\delta G_k\, \rho\,
\mathrm{diag}(1,1,1,1)\Big]\label{quantumeinstein}
\end{align}
\end{subequations}
In (\ref{quantumeinstein}) the Ricci tensor and its trace are
understood to be computed from $h_{\mu\nu}^c$ exclusively. The
classical Newtonian potential $\phi(\mathbf{x})$ is equal to
$-\frac{1}{2}h_{00}^c$ and, per (\ref{poisson}), is found solving
Poisson's equation as usual. Its quantum correction, bearing the
same relation to $h_{00}^q$, will be found from
\begin{equation}\label{h00nonloc}
h^q_{00}(\mathbf{x})=\delta G_k h_{00}^c(\mathbf{x})- 256 \pi^2
G_0^2 \left[F_{1,k}(-\nabla^2)+F_{2,k}(-\nabla^2)\right]
\rho(\mathbf{x})\,,
\end{equation}
which is obtained replacing in (\ref{quantumeinstein})
$R_{\mu\nu}$ by $-\frac{1}{2}\nabla^2 h_{\mu\nu}^c$, using
(\ref{poisson}), and cancelling Laplacians. Therefore, the quantum
correction to the Newtonian potential consists of two terms: a
trivial shift due to renormalization of $G$, plus a nontrivial
part that is found by direct application of the nonlocal form
factor $F_1+F_2$ to the classical matter distribution. From now on
we take $h_{00}^q$ to refer only to the nontrivial part, absorbing
the first term in a redefinition of $h_{00}^c$.

Since the Laplacian in all the preceding expressions is flat (to
keep only the first order in the metric perturbation), the action
of the form factors in (\ref{h00nonloc}) can be computed with an
ordinary Fourier transform:
\begin{align}\label{fourier}
h^q_{00}(\mathbf{x})&=-256 \pi^2 G_0^2 \frac{1}{(2\pi)^3}\int\mathrm{d}^3x'\rho(\mathbf{x}')\int\mathrm{d}^3p\,\left[F_{1,k}(p^2)+F_{2,k}(p^2)\right] \mathrm{e}^{-i \mathbf{p}\cdot(\mathbf{x}-\mathbf{x}')} \nonumber\\
&=-\frac{4G_0^2}{\pi^2}\int \mathrm{d}^3 x'
\frac{\rho(\mathbf{x}')}{\left|\mathbf{x}-\mathbf{x}'\right|}\int_0^\infty\mathrm{d}p\,\,p
F_k(p) \sin(p \left|\mathbf{x}-\mathbf{x}'\right|)\,\,,
\end{align}
where we have defined
\begin{align}\label{defFp}
F_k(p)&=32\pi^2(F_{1,k}(p^2)+F_{2,k}(p^2))\nonumber\\
&=\frac{43}{60}\log\left(\frac{p^2+(k^2-p^2)\theta(4k^2-p^2)}{k_0^2}\right)+\Bigg[\left(-\frac{1093}{900}-\frac{373}{450}\frac{k^2}{p^2}-\frac{14}{75}\frac{k^4}{p^4}\right)\sqrt{1-\frac{4k^2}{p^2}}\nonumber\\
&\,\,\,\,+\frac{43}{30}\log\left(\frac{1+\sqrt{1-\frac{4k^2}{p^2}}}{2}\right)\Bigg]\theta(p^2-4k^2)
\end{align}
Equations (\ref{fourier}) and (\ref{defFp}) comprise the result we
wanted for the quantum correction. For further analysis we call
$I_1$ the term of the $p$-integral in (\ref{fourier}) that comes
from the first term of (\ref{defFp}), and $I_2$ the $p$-integral
of the remaining terms. $I_1$ can be evaluated exactly as a
combination of elementary integrals and distributional Fourier
transforms, whereas in $I_2$ no such closed form can be found. We
have for $I_1$:
\begin{align}
I_1&=\frac{43}{60}\left[\log\left(\frac{k^2}{k_0^2}\right)\int_0^{2k}\mathrm{d}p\,p\,\sin(p\left|\mathbf{x}-\mathbf{x}'\right|)+  \int_{2k}^\infty \mathrm{d}p\,p\,\log\left(\frac{p^2}{k_0^2}\right)\sin(p\left|\mathbf{x}-\mathbf{x}'\right|)\right]\,\nonumber\\
&=\frac{43}{60}\left[\int_0^\infty\mathrm{d}p\,p\,\log\left(\frac{p^2}{k_0^2}\right)\sin(p\left|\mathbf{x}-\mathbf{x}'\right|)\,-\int_0^{2k}
\mathrm{d}p\,p\,\log\left(\frac{p^2}{k^2}\right)\sin(p\left|\mathbf{x}-\mathbf{x}'\right|)\right]
\end{align}
The first integral, though divergent, is well-defined as the
Fourier transform of a generalized function and can be computed
from the formulas in \cite{lighthill}; the second one can be computed
analytically. Thus:
\begin{align}
I_1&=\frac{43}{60}\Bigg[-\frac{\pi}{X^2}+\log(k_0^2)\frac{\delta'(X)}{2}
+\log(4)\left(\frac{\sin(2kX)}{X^2}-\frac{2k\cos(2kX)}{X}\right)\nonumber\\
&+\frac{2\mathrm{Si}(2kX)}{X^2}-\frac{2\sin(2kX)}{X^2}\Bigg]\,,
\end{align}
where we defined $X=\left|\mathbf{x}-\mathbf{x}'\right|$.

The remaining terms, comprising $I_2$, can also be rewritten as a
combination of distributional Fourier transforms and convergent
integrals, but for them the convergent part cannot usually be
computed in closed form (though it can be investigated numerically
if so desired). For this reason, we restrict ourselves to
evaluating the large $X$ asymptotic expansion of the result. If
the matter distribution is a point source, i.e.
$\rho(\mathbf{x})=M\delta^3(\mathbf{x})$, this will give us the
long-distance quantum corrections to the Newtonian potential of a
point source in an asymptotic series. However, note that we will
later call into question the physical validity of such an
asymptotic expansion, so the following calculation needs to be
taken with a grain of salt.

The large frequency asymptotic form of the Fourier transform of a
generalized function of the kind we have can be found using
Theorem 19 from reference \cite{lighthill}. Applying it gives us that
for large $X$ we have:
\begin{align}
I_2&=-\frac{43}{30}\log(4)\frac{k\cos(2kX)}{X}-\frac{7}{200}\sqrt{\pi k
}\frac{\sin\left(2kX+\frac{3\pi}{4}\right)}{X^{3/2}}\nonumber\\
&+\frac{43}{30}\left(1+\log 2\right)\frac{\sin(2kX)}{X^2}+o\left(\frac{1}{X^2}\right)\,.
\end{align}
Joining this with the asymptotic expansion for the result we
obtained for $I_1$, we conclude that the quantum correction for
the Newtonian potential of a point source of mass $M$ is given at
long distances $r=|\mathbf{x}|$ by
\begin{equation}
h_{00}^q(r)=-\frac{8MG_0^2k}{\pi
r^2}\left[\frac{43}{60}-\frac{7}{400\sqrt{\pi
}}\frac{\sin\left(2kr+3\pi/4\right)}{\sqrt{kr}}-\frac{43}{120\pi}\frac{\left(2\cos(2kr)+\pi\right)}{kr}+o\left(\frac{1}{k
r}\right)\right]\,.
\end{equation}
The full Newtonian potential of the point source would therefore be
given asymptotically by:
\begin{align}\label{phiasympt}
\phi(r)&= -\frac{MG_0}{r}\Bigg[1+\delta G_k-\frac{43G_0 k}{15\pi r}+\frac{7G_0\sqrt{k}}{100\pi^{3/2 }}\frac{\sin\left(2kr+3\pi/4\right)}{r^{3/2}}\nonumber\\
&+\frac{43G_0}{30\pi^2}\frac{\left(2\cos(2kr)+\pi\right)}{r^2}+o\left(\frac{1}{r^2}\right)\Bigg]\,.
\end{align}

One might find this asymptotic result peculiar in several
respects: the leading order correction is dominant over the EFT
ones ($k/r^2$ versus $1/r^3$), and the subleading terms are
oscillatory, which is difficult to interpret physically. We suggest that this is due to the  large $X$ expansion being unphysical for the problem under consideration. Recall that the equations of motion derived from
the effective action predict the expectation values of quantities
such as the metric. From the effective average action, which is
defined by a path integral with an infrared cutoff at scale $k$,
we would expect to obtain, if $k$ is treated as fixed, equations of motion for ``approximate
expectation values'', that do not include the information about
low frequency field modes. These might be a good approximation in
the $kr\ll 1$ regime (where $r$ is the radial coordinate for us,
or more in general the greatest physical length scale involved in
the problem) but we would expect the results to be incorrect for
$kr\gg 1$. If this reasoning is accepted, the result in
(\ref{phiasympt}) is not physical, since it is obtained as a large
$r$ expansion for fixed $k$; the physical quantum correction to
the potential is to be obtained instead from the $k\rightarrow 0$
limit at fixed $r$.

Taking the limit $k\rightarrow 0$ in (\ref{fourier}) in $I_2$
gives zero up to a purely local term involving $\delta'(r)$. In
$I_1$, this results in a similar local term plus $43\pi/(60r^2)$.
This latter term implies a nontrivial long-distance correction for
the Newtonian potential
\begin{equation}\label{dono}
\phi(r)= -\frac{MG_0}{r}\left[1+\frac{43G_0}{30\pi r^2}\right]
\end{equation}
(because $\delta G_k$, obtained by integrating the flow
(\ref{betas}), is $\sim G_0k^2$ and thus vanishes at $k=0$). This
is the same result obtained in effective quantum gravity for the
contribution of the graviton and ghost vacuum polarization
diagrams to $\phi$ \cite{Donoghue:1994dn, Dalvit:1997yc}. Since we have treated the matter source
in a purely classical way, we are not obtaining the terms due to
vertex corrections\footnote{A local term with $\delta'(r)$ is also
found in effective quantum gravity, with an arbitrary constant
involving the renormalization scale $\mu$.}. This confirms that
the effective field theory quantum corrections to Newton's law are
indeed recovered in the right limit from the effective average
action.

Our remarks above imply a general philosophy of interpreting the
effective average action as a useful device to follow the flow of
the renormalization group (and hopefully discover a UV fixed
point) but not as an action from which physics can be directly
extracted by solving equations of motion: these, predicting expectation values of observables,
should be computed from the $k=0$ usual effective action. On the
other hand, there have been several attempts \cite{Bonanno:2000ep,Reuter:2003ca,Reuter:2004nv,Falls:2010he} to extract
physics from the effective average action itself by identifying
$k$ with an inverse distance scale of the spacetime under
consideration, instead of as a fixed parameter. For a static and
spherically symmetric situation like the one we are considering
(with just a point source) this proposal means $k=\zeta/r$ with
$\zeta$ a numerical constant. (Note that above we argued that the
effective average action with cutoff $k$ is to be trusted only for
$k\ll r^{-1}$, whereas this approach conjectures that it can be
trusted only for $k=\zeta r^{-1}$.). If we do this replacement
before computing $I_1$ and $I_2$, we see that in the case of a
point source the variable $x$ is just $2\zeta$, and so all the
functions of it are numerical constants. The Newtonian potential
would therefore be
\begin{equation}
\phi(r)= -\frac{MG_0}{r}\left[1+ \frac{\xi\,G_0}{r^2}\right]
\end{equation}
with $\xi$ a numerical constant (the $\delta G_k$ term gives a
similar contribution, as has often been noted). This agrees with
the general form of the effective field theory correction
discussed above. However, this agreement in form is rather
trivial, following just from dimensional analysis and the lack of
other length scales in the problem once we make the conjectured
identification for $k$. A more interesting test of the conjectured
identification is in the case of an extended source with spherical
symmetry, such as a spherical star. Here a naive application of
this conjecture that equates $k$ with the inverse distance to the
center of the star, at the level of (\ref{fourier}), leads to
\begin{equation}\label{identified}
h^q_{00}(\mathbf{x})=-\frac{4\zeta^2G_0^2}{\pi^2}\int \mathrm{d}^3
x'
\frac{\rho(\mathbf{x}')}{\left|\mathbf{x}-\mathbf{x}'\right|r^2}f\left(\frac{\zeta\left|\mathbf{x}-\mathbf{x}'\right|}{r}\right)\,,
\end{equation}
which is obtained from (\ref{fourier}) by making the replacement $k=\zeta/r$, changing variables from $p$ to $v=p r$, and defining $f$ as the result of the $v$-integrals $I_1+I_2$ with
$1/r^2$ taken out. On the other hand, the result obtained from
effective theory (and from our formulas as $k\rightarrow 0$) is
\begin{equation}\label{star}
h^q_{00}(\mathbf{x})=\frac{43G_0^2}{15\pi}\int \mathrm{d}^3 x'
\frac{\rho(\mathbf{x}')}{\left|\mathbf{x}-\mathbf{x}'\right|^3}\,.
\end{equation}
These two formulas agree asymptotically for large $r$ (up to a
factor related to $\zeta$), but will in general be different at
closer distances where results depend on new length scales
implicit in the function $\rho(r)$ for the density within the
star (see \cite{Satz:2004hf} for a discussion of how
(\ref{star}) depends on this).

It is not surprising that the direct identification $k=\zeta/r$ fails, even in spherically symmetric spacetimes, when there is more than one length scale of interest. $k$ is a momentum scale, and matching it with a spacetime scale, as its inverse, should be possible unambiguously when there is only one spacetime scale to be used (such as $r$ for a point source). But when there are more length scales involved, such as the radius of a star or other parameters relevant to its internal structure, there is no a priori reason to expect the simple matching of $k$ with inverse radial distance to be correct. It is noteworthy that the qualitative agreement with effective field theory for a general source could be obtained if (instead of $r^{-1}$) we identified $k$ with $\left|\mathbf{x}-\mathbf{x}'\right|^{-1}$ before integrating over $\mathbf{x}'$, in effect defining a $k(\mathbf{x},\mathbf{x}')$ as the inverse distance to each point source $\mathbf{x}'$ instead of the inverse distance to the global center of symmetry of the spacetime. The plausible reason why this works is because for each point source the direct identification with inverse radial distance is valid, and within our weak field approximation linear superposition of sources is valid; we make no claim as to its validity in a more general situation, a matter which deserves further research. The alternative position that we are suggesting is that the equations of motion derived from the effective average action are to be trusted only in a small $k$ regime, and ultimately are only correct when $k\rightarrow 0$ and we recover the usual effective action.

\section*{Conclusions}

The effective average action and its renormalization flow are
promising research avenues for developing a theory of quantum
gravity. In this paper we have sought to accomplish two related
goals. The first one is to take the first steps towards
investigating the flow of non-local terms in the action. We have
been able to compute and exhibit the solution for the non-local
form factors in the second-order term in a general curvature
expansion for the action, within the one-loop approximation 
where the flow is generated only by the operator $\int\sqrt{g}R$
Equations (\ref{20.1}) and (\ref{20.2}) summarize this result.

It would obviously be desirable to go beyond our approximations.
One-loop computations of the flow of higher order non-local terms
seems possible in principle by extension of the methods of this
paper, though computationally very intensive. On the other hand,
going beyond the one-loop approximation and obtaining information
about the non-perturbative flow is a difficult and open research
question. A first step towards it would be to improve the running
of the form factors (\ref{14}) by computing the beta function not
from the bare action but from an action with running $G_{k}$ and
$\Lambda_{k}$, which can be found from the beta functions
(\ref{betas}).

 Another question is how our results generalize to $d$ dimensions; so far, the only
similar case that has been investigated is the $d=2$ computation
for the Polyakov action in \cite{Codello:2010mj}. Local truncations in extra dimensions were explored in \cite{fischerlitim}.

Our second goal has been to clarify the relationship between the
effective average action formalism and the usual perturbative
effective field theory. These goals are related in that the
one-loop effective action for quantum gravity is non-local. We have
shown how this action is recovered in the $k\rightarrow 0$ limit
of our flow, and how the long-distance quantum corrections it
induces on the Newtonian potential are also recovered in the same
limit. We have also exhibited a computation of the quantum
correction from the $k$-dependent average action, and we have
argued that it cannot be trusted except in the $k\rightarrow 0$
limit, and in particular not in the $kr\gg 1$ regime. It has been
shown in passing that the often-suggested prescription of
identifying $k$ with the inverse radial distance does not recover
effective theory results in general, beyond the simplest case of a
point source. However, more sophisticated relationships of $k$ with inverse distance scales (such as the one considered in the last paragraph of the previous section) are not ruled out and require further research. Our results might also be of interest in applications of the effective average action to cosmology, where similar identifications of $k$ with inverse time scales have been suggested \cite{Bonanno:2007wg,Bonanno:2010bt}.

A central question faced by the effective average action formalism
is how to extract physical predictions from the flow. In our
opinion, our results lend credence to the position that the
relevant physics is to be extracted from the full effective action
at $k\rightarrow 0$. $\Gamma_k$ would then be considered a useful
device for studying the flow of the theory and unveiling a UV
fixed point, but not to compute observables, which should in
principle be computed (at all scales) from the effective action
$\Gamma_0$. This is to be taken only as a tentative position, and more work is needed on the foundational conceptual
issues of the effective average action approach to settle these
questions definitely.

\appendix

\section{Non-local heat kernel expansion and computation of the beta functions}

Our problem is to compute functional traces of the form $\mathrm{Tr}[W(\Delta)]$
where $W$ for gravitons and for ghosts is given in the two terms
of (\ref{3}) and $\Delta$ for each is given in (\ref{7}). Following
standard practice, e.g. \cite{Codello:2008vh}, we use $K(s)$, the heat kernel
of the operator $\Delta$: \begin{equation}
\mathrm{Tr}[W(\Delta)]=\int_{0}^{\infty}\mathrm{d}s\,\mathrm{Tr}[K(s)]\tilde{W}(s)\end{equation}
 where $\tilde{W}$ is the Laplace anti-transform of $W$. A general
expression for the trace of the heat kernel of operators of the form
(\ref{7}), in the form of a curvature expansion with non-local form
factors, is obtained in \cite{Barvinsky:1990up,Avramidi:1990je}. Using a notation
in which $\Delta=-\square-\hat{P}+\frac{1}{6}R\hat{1}$, it reads
\begin{align}\label{kernel}
\mathrm{Tr}[K(s)] & =\frac{1}{(4\pi s)^{d/2}} \int \mathrm{d}^{d}x\,\sqrt{-g}\,\mathrm{tr}\Bigg\{\hat{1}+s\hat{P}+s^{2}\Big[R_{\mu\nu}\phi_{1}(-s\square)R^{\mu\nu}\hat{1}+R\phi_{2}(-s\square)R\hat{1}\nonumber \\
&+\hat{P}\phi_{3}(-s\square)R+\hat{P}\phi_{4}(-s\square)\hat{P}+\mathcal{\hat{R}}_{\mu\nu}\phi_{5}(-s\square)\mathcal{\hat{R}}^{\mu\nu}\Big]\Bigg\}+O(\mathcal{R}^{3})\,,\end{align}
 where $\mathcal{\hat{R}}_{\mu\nu}$ is the curvature of the covariant
derivative acting on the space where the operator $\Delta$ acts.
In our case we have $\mathcal{\hat{R}}_{\mu\nu}=-2R_{\,\,\,(\alpha\mu\nu}^{(\gamma}\delta_{\,\beta)}^{\sigma)}$
for the graviton case, and $\mathcal{\hat{R}}_{\mu\nu}=R_{\,\,\beta\mu\nu}^{\alpha}$
in the ghost case. The functions $\phi_{i}(x)$ take the form: \begin{subequations}
\begin{align}
\phi_{1}(x) & =\frac{f(x)-1+\frac{1}{6}x}{x^{2}}\,,\\
\phi_{2}(x) & =\frac{1}{8}\left[\frac{1}{36}f(x)+\frac{f(x)-1}{3x}-\frac{f(x)-1+\frac{1}{6}x}{x^{2}}\right]\,,\\
\phi_{3}(x) & =\frac{1}{12}f(x)+\frac{f(x)-1}{x}\,,\\
\phi_{4}(x) & =\frac{1}{2}f(x)\,,\\
\phi_{5}(x) & =-\frac{1}{2}\frac{f(x)-1}{x}\,,\end{align}
 \end{subequations} where $f(x)$ is defined above in (\ref{fdef}).
To obtain our claimed result (\ref{8}), all we need to do is reexpress
the five second-order terms in (\ref{kernel}) in terms of the basis
$R_{\mu\nu}R^{\mu\nu}$, $R^{2}$. This is done by expressing $\mathcal{\hat{R}}_{\mu\nu}$
in terms of $R_{\alpha\beta\mu\nu}$ as defined above, expressing
$\hat{P}$ in terms of $R_{\alpha\beta\mu\nu}$, $R_{\mu\nu}$ and
$R$ by using the definition (\ref{4}) for the $\Delta$ operators,
and eliminating $R_{\alpha\beta\mu\nu}$ in terms of $R_{\mu\nu}$
and $R$ by means of the identity \begin{equation}
R_{\alpha\beta\mu\nu}=\nabla_{\mu}\nabla_{\alpha}\frac{1}{\square}R_{\nu\beta}-\nabla_{\nu}\nabla_{\alpha}\frac{1}{\square}R_{\mu\beta}-\nabla_{\mu}\nabla_{\beta}\frac{1}{\square}R_{\nu\alpha}+\nabla_{\nu}\nabla_{\beta}\frac{1}{\square}R_{\mu\alpha}+O(\mathcal{R}^{2}),\end{equation}
 proven in \cite{Barvinsky:1990up}. Note that to the order
we are working we can freely commute covariant derivates as well as
functions of them, throwing away terms of higher order in the curvature
arising from commutators. Doing the described operations for the graviton
part and the ghost part of the trace and combining them we obtain
result (\ref{8}).

\section{The $Q$-functionals and their integrals}

The $Q$-functionals of a function $f(z)$ are defined by:
\begin{equation}
Q_{n}[f]=\int_{0}^{\infty}\mathrm{d}s\,\tilde{f}(s)s^{-n}\,,\label{Q_0}
\end{equation}
where $\tilde{f}(s)$ is the anti-Laplace transform of $f(z)$. They have an equivalent expression in terms of $f(z)$ as:
\begin{equation}
Q_{n}[f]=\left\{ \begin{array}{ccc}
\frac{1}{\Gamma(n)}\int_{0}^{\infty}\mathrm{d}z\,z^{n-1}f(z) &  & n>0\\
(-1)f^{(n)}(0) &  & n\leq0\end{array}\right..\label{Q_01}
\end{equation}
In the case an additional factor $e^{-as}$ is present in (\ref{Q_0}) the functions $f(z)$ in (\ref{Q_01}) are to be evaluated at the shifted point $z+a$.

Our expressions for the traces can be rewritten in terms of $Q$-functionals
with $n=0,1,2$. We encounter $Q$-functionals inside parameter
integrals of the form:
\begin{equation} \int_{0}^{1}\mathrm{d}\xi\,
Q_{n}\left[h_{k}\left(z+x\xi(1-\xi)\right)\right]\,,\label{Q_1}
\end{equation}
where recall that $h_{k}(z)=\frac{\partial_{t}R_{k}(z)}{z+R_{k}(z)}$.
 These integrals can be calculated analytically if we employ the optimized
cutoff (\ref{optim}). The results are: 
\begin{equation}
\int_{0}^{1}\mathrm{d}\xi\,
Q_{0}\left[h_{k}\left(z+x\xi(1-\xi)\right)\right]=2\left[1-\sqrt{1-\frac{4}{u}}\theta(u-4)\right]\,,\label{Q_6}
\end{equation}
\begin{equation}
\int_{0}^{1}\mathrm{d}\xi\,
Q_{1}\left[h_{k}\left(z+x\xi(1-\xi)\right)\right]=2k^{2}\left[1-\frac{u}{6}+\frac{u}{6}\left(1-\frac{4}{u}\right)^{\frac{3}{2}}\theta(u-4)\right]\,,\label{Q_7}
\end{equation}
\begin{equation}
\int_{0}^{1}\mathrm{d}\xi\,
Q_{2}\left[h_{k}\left(z+x\xi(1-\xi)\right)\right]=2k^{4}\left[\frac{1}{2}-\frac{u}{6}+\frac{u^{2}}{60}-\frac{u^{2}}{60}\left(1-\frac{4}{u}\right)^{\frac{5}{2}}\theta(u-4)\right]\,,\label{Q_8}
\end{equation}
where $u=x/k^2$. Using these results in (\ref{13}) leads to the beta functions (\ref{15}) and (\ref{16}).
\\
\bigskip
\\
\noindent{{\large\bf Acknowledgements}}\\
\\
A.S. would like to thank the Martin Reuter and the Theoretical Physics research group at the Mainz University for generous hospitality. We thank Martin Reuter, Frank Sauressig and Elisa Manrique for helpful discussions. The work of A.S. and F.D.M. was supported by UBA, CONICET\ and ANPCyT.

\end{document}